\documentclass[11pt,twoside]{article}
\usepackage{asp2004}
\usepackage{psfig}
\usepackage{epsf}
\usepackage{graphics}
\usepackage{lscape}
\def\edcomment#1{\iffalse\marginpar{\raggedright\sl#1\/}\else\relax\fi}
\reversemarginpar

\begin{document}
\title{Observations of Outflows in Cataclysmic Variables}
\author{Cynthia S. Froning}
\affil{Center for Astrophysics and Space Astronomy, University of
Colorado, 593 UCB, Boulder, CO 80309, USA}

\begin{abstract}
Outflows in the form of fast winds from the accretion disk appear
in nearly all high accretion rate cataclysmic variables (CVs),
both novalikes (NLs) and dwarf novae in outburst (DN). The primary
signatures of CV winds are broad, blueshifted absorption features
and P~Cygni profiles that appear in the ionized metal lines of
lower inclination systems and the broad emission lines that appear
in eclipsing systems. Much progress has been made in our
understanding of the structure and behavior of CV outflows based
on data obtained with X-ray, EUV, and UV telescopes and the
kinematical models developed to fit the observations.  Our current
picture of CV outflows is that they are bipolar, rotating winds
driven off the accretion disk primarily by radiative line driving
from disk, boundary layer, and white dwarf radiation.  Other
mechanical forces may play a role in the formation of CV winds.
The winds are highly turbulent and show complex vertical and
azimuthal structure. Future work will focus on determining the
physical, thermal and ionization structure of the winds and
probing the mechanisms that govern wind formation and behavior.
\end{abstract}

\thispagestyle{plain}

\section{Background}

\subsection{The First Observations of CV Winds}

The study of outflows in cataclysmic variables (CVs) was born in
the IUE era.  IUE ultraviolet spectra of luminous CVs ---
novalikes (NLs) and dwarf novae (DN) in outburst --- showed high
ionization UV transitions, particularly of CIV
$\lambda\lambda$1548,1552~\AA, SiIV $\lambda\lambda$1393,1402~\AA\
and NV $\lambda\lambda$1238,1242~\AA, appearing with very broad,
blueshifted absorption components and P Cygni profiles, shifting
to pure emission in high orbital inclination systems
\citep{heap1978,krautter1981,klare1982,greenstein1982,szkody1982,cordova1982,hassall1983}.
Examples of wind-dominated FUV spectra of CVs at different
inclinations are given in Figure~1, which shows observations taken
from the Far Ultraviolet Spectroscopic Explorer (FUSE) data
archive. It was immediately noted that the line profiles were
similar to those seen in early-type and luminous stars such as O
stars and Wolf-Rayet stars, in which the lines were (and are)
believed to originate in expanding winds driven by radiation
pressure \citep{conti1978}. As a result, a wind origin for the UV
lines of high accretion rate CVs was adopted early on.

\begin{figure}[!ht]
\plotone{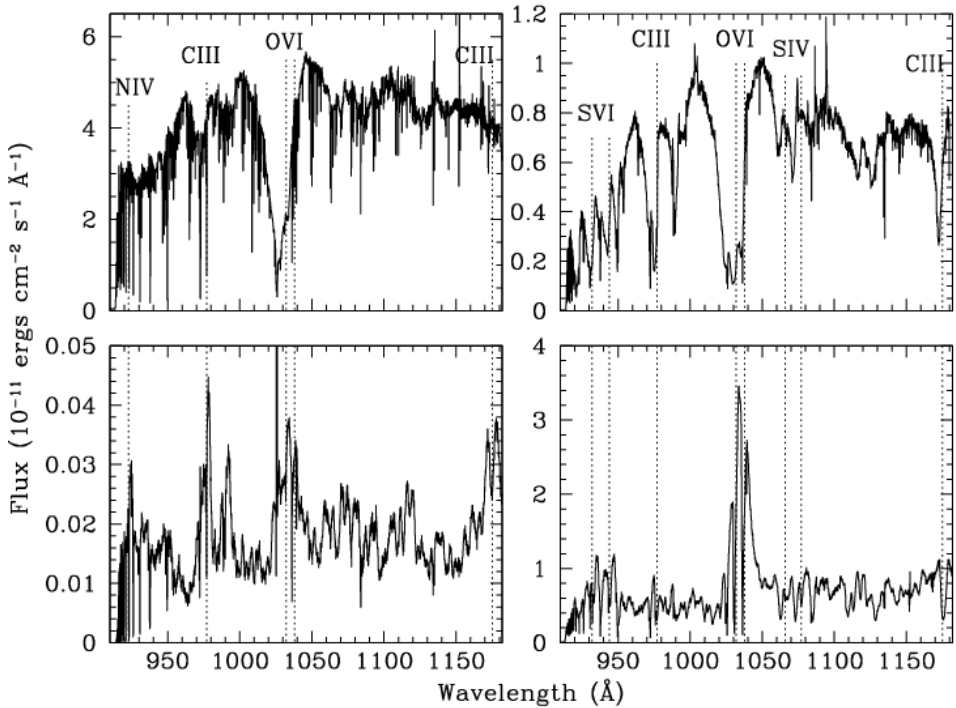} \caption{Examples of CV FUV spectra for
different system inclination angles.  The spectra are
time-averaged plots of observations in the FUSE data archive
binned to 0.1~\AA. The upper left panel shows the spectrum of the
DN SS~Cyg in outburst ($i = 30^{\circ}$).  The upper right panel
shows the spectrum of the NL IX~Vel ($i = 60^{\circ}$). The lower
left panel shows the spectrum of the NL, UX~UMa ($i =
71^{\circ}$), and the lower right panel shows the spectrum of the
DN, WZ~Sge, at outburst peak ($i = 75^{\circ}$). Analysis of the
UX~UMa and WZ~Sge observations can be found in Froning et al.\
2003 and Long et al.\ 2003, respectively.}
\end{figure}

IUE spectroscopy revealed several properties of the CV winds. The
maximum blue edge velocities of the absorption troughs
($\simeq5000$~km~s$^{-1}$), which give a lower limit to the
terminal velocity of the wind, were of order the escape velocity
from the white dwarf (WD) accretor, suggesting that the outflows
originate close to the WD \citep{cordova1982}. The deepest
absorption in the lines occured not at the blue edges of the
lines, as is the case for early-type stars, but at lower
velocities ($\simeq2000$~km~s$^{-1}$), suggesting that the wind is
more slowly accelerated in CVs than in luminous stars
\citep{mauche1987}. In eclipsing systems, the lines, unlike the
continuum, were not occulted, indicating that the line-emitting
region was extended relative to the bulk of the disk, with sizes
of order the size of the occulting donor star
\citep{holm1982,mauche1994}.  A comparison of the strengths of the
lines with model lines from spherically-symmetric, radiation
pressure driven winds (designed for luminous stars) gave mass-loss
rates in the wind from $10^{-11}$ -- $10^{-12}
M_{\odot}$~yr$^{-1}$, or $10^{-2}$ -- $10^{-4}$ of the mass
accretion rate in the disk
\citep{krautter1981,klare1982,greenstein1982,cordova1982}.

\subsection{Early CV Wind Models}

From the start, models for CV winds have been based on the wind
models for luminous stars:  resonant line scattering from
absorption of photons from the central source transfers momentum
to ions; as the wind expands outwards, the local resonance is
redshifted, perpetuating the line scattering and the driving of
the outflow. (For a more detailed discussion of the physics of CV
wind models, see Proga, this volume.) The first models assumed a
radial outflow with a constant ionization structure emerging from
the center of the disk and the WD
\citep{drew1985,drew1987,mauche1987}. The models were used to
compare theoretical line profiles to the observed lines, in
particular CIV, the strongest wind feature in the UV spectrum.

The results revealed problems with the picture of a radial wind.
In high inclination CVs, the blueshifted absorption component of
the wind lines disappears as the disk becomes more edge-on, but
the radial models continue to show absorption at high inclination.
\citet{drew1987} cited this as evidence for bipolarity, rather
than spherical symmetry, in the outflow, although it was noted by
\citet{mauche1987} that radial winds could still be present if the
bipolarity was introduced by limb-darkening effects in the
accretion disk. Mauche \& Raymond pointed out another significant
problem with radial CV winds, however: radial winds ``see'' so
much of the continuum from the WD and boundary layer (BL) that the
wind is ionized beyond CIV unless it is highly optically thick,
which requires wind mass loss rates of $\geq$1/3 of the mass
accretion rate. Such a massive wind cannot be radiatively driven,
produces UV line fluxes an order of magnitude larger than observed
values, and would absorb the soft X-ray continuum from the BL,
which is inconsistent with observed X-ray fluxes. On the basis of
their models, the authors concluded that radial outflows were
unlikely in CVs.

As a result, the next generation of CV wind models adopted
biconical, axisymmetric wind structures, with the winds being
launched from the accretion disk. \citet{shlosman1993} developed a
kinematical model of a rotating, biconical disk wind, taking into
account radiation from the WD, the BL, and the accretion disk when
calculating the local ionization structure of the (constant
temperature) wind. Radiation transfer and line driving of the wind
were treated using the standard wind modeling theory for luminous
stars. By comparing theoretical line profiles from their models
with those of radial winds, they showed that biconical disk winds
provide a better match to the profiles of CIV wind lines at
varying inclinations and do not suffer from the excessive
absorption of BL photons and subsequent over-ionization of the
wind as seen in radial wind models.   \citet{vitello1993} compared
the biconical disk model line profiles to observed IUE spectra of
CVs. They were able to match the CIV profiles of the low
inclination system, RW~Sex, and the eclipsing NL, RW~Tri, with
wind mass-loss rates of order 1 -- 15\% of the mass accretion
rate. Their models gave a typical scale for the CIV line of 50 --
100~R$_{WD}$ in vertical height above the accretion disk. At this
point, the theoretical and observational evidence both pointed to
the presence of biconical accretion disk winds driven by radiation
pressure in high accretion rate CVs.

\section{Properties of Cataclysmic Variable Outflows}

\subsection{Wind Structure and Behavior}

Much of the progress in understanding CV outflows continues to
come from UV spectroscopic observations of luminous systems,
although EUV and X-ray observations of CVs have also provided
information on the properties of the winds.  Winds have also been
cited as possible sources for features in optical lines,
particularly non-Keplerian emission in HeII $\lambda4686$~\AA, but
the evidence for wind signatures at optical wavelengths is
generally indirect \citep{honeycutt1986,marsh1990}. One exception
is the optical spectrum of BZ~Cam, which shows intermittent
P~Cygni profiles in H$\alpha$ and HeI $\lambda$5876~\AA\
\citep{patterson1996,ringwald1998}. The optical lines are complex
blends of emission from the wind and the disk, however, which
limits analysis of the wind behavior based on optical emission. In
the UV, the advent of sensitive, high time resolution UV
spectrographs on HST and FUSE has allowed direct tests of the
predictions of disk wind models. Below, we examine several
properties of CV winds that have resulted from observations in the
past decade.

\subsubsection{Wind Rotation}

One observational signature of line-driven disk winds is that they
rotate and that the winds preserve angular momentum as they leave
the disk. This is in contrast to hydromagnetically-driven winds,
such as those seen in YSOs, which maintain a constant angular
velocity out to the Alfv\'{e}n radius and perhaps beyond.  In at
least two CVs, the observational evidence favors a rotating wind
that conserves angular momentum.  \citet{shlosman1996} compared
the out of eclipse and mid-eclipse UV line profiles in IUE
observations of the nearly edge-on NL, V347~Pup.  They showed that
the CIV wind line narrows in eclipse when the region of the wind
near the disk surface that is responsible for the most of the
rotational broadening of the line is occulted.  They demonstrated
that the line profile variations during eclipse are consistent
with the biconical disk wind model of \citet{shlosman1993} when
only orbital phase changes are invoked.  The CIV emission line
also narrows during eclipse in UX~UMa and a rotational signature
can be seen in the eclipse light curves of different velocities in
the line:  the eclipse appears in the blue wing of the line before
it appears in the red wing, and the eclipse is most shallow at the
line center \citep{mason1995,baptista1995}. \citet{knigge1997}
modeled the UX~UMa data --- high spectral resolution,
time-resolved HST/GHRS observations of CIV --- with a biconical
disk wind model and were able to reproduce the pre- and
mid-eclipse spectra and the velocity-resolved eclipse light curves
with moderately collimated (65$^{\circ}$ opening angle), rotating
disk wind.

\subsubsection{Wind Vertical Structure}

Observations of eclipsing systems have also been instrumental in
probing the vertical structure of the winds as they are launched
from the disk.  The HST observations of the CIV line in UX~UMa, in
addition to exhibiting a rotational disturbance, show narrow,
low-velocity absorption components that are occulted during
eclipse.  The same narrow absorption components are observed in
numerous other lines in the HST spectral range and also appear in
phase-resolved FUSE spectra of UX~UMa \citep{froning2003}.  The
eclipse behavior of the UV spectrum of UX~UMa is shown in
Figure~2.  Outside of eclipse, the spectrum is a complex blend of
line emission, with no discernable continuum. At mid-eclipse, the
spectrum is much cleaner, and is characterized by a flat continuum
and strong, broad line emission from the unocculted portion of the
wind.  The difference spectrum, which is the spectrum of the
eclipsed light, shows that the eclipsed parts of the wind lines
are narrow ($\simeq500$ -- 1000~km~s$^{-1}$) absorption components
centered on the rest wavelengths of the lines. \citet{knigge1997}
showed that to successfully model these absorption dips and their
disappearance during eclipse requires the presence of a dense, low
outflow velocity transition region between the bulk of the
accretion disk and the fast wind.  The transition region, or disk
chromosphere, may be part of the wind itself:  the hydrodynamical
disk wind models of \citet{proga1998} show that when the line
driving is dominated by radiation from the disk (over that from
the WD and BL), the wind has both a fast outflow component and a
dense, low outflow velocity region near the disk surface.

\begin{figure}[!ht]
\plotone{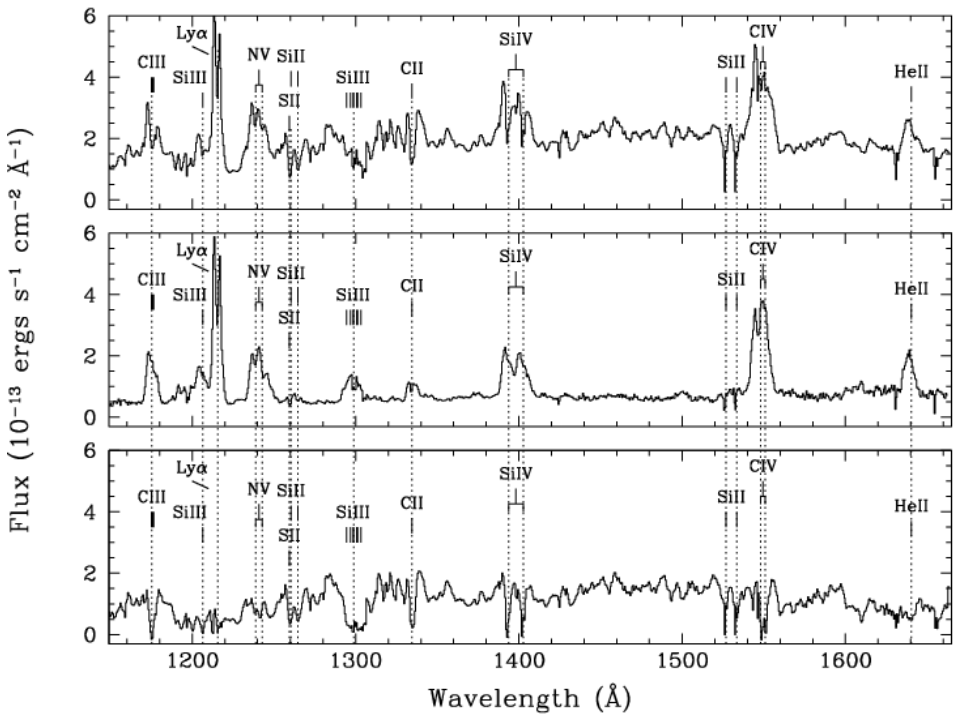}\caption{HST/GHRS FUV spectra of UX UMa.
The upper panel shows the out of eclipse spectrum of UX UMa. The
middle panel shows the spectrum at mid-eclipse, and the lower
panel shows the difference spectrum, which is the spectrum of the
eclipsed light.  Dotted lines indicate the rest wavelengths of the
observed spectral features. (Figure from Froning et al.\ 2003. See
also Mason, Drew, \& Knigge 1997 for a discussion of pre-eclipse
light curve dips in these observations.)}
\end{figure}

\subsubsection{Orbital Variability}

Recent observations also point to azimuthal structural variations
in the disk winds. Time-resolved FUSE spectra of RW~Sex show that
the FUV wind lines are modulated on the orbital period
\citep{prinja2003}. Over the orbit, all of the wind lines in the
FUV spectrum --- ranging from lower-ionization transitions of CIII
and NIII to PV, SVI, and OVI --- show the same behavior: a
single-humped modulation in which the velocity of the deepest
absorption of each line moves to the red, accompanied by an
increase in the absorption depth, and a shift back. The authors
speculate on the source of the orbital phase modulation, which
they link to an asymmetry in the structure of the disk.
Orbital-phased modulations in wind lines have also been observed
in the NLs V795~Her and V592~Cas \citep{rosen1998,prinja2004}.

The source of the orbital changes in the wind lines remains
unknown.  \citet{prinja2003} suggest that a disk tilt or warp
could account for the modulation in RW~Sex, but this explanation
is not favored for V592~Cas, as the wind variations are modulated
on the orbital period, but not on the known superhump period of
the system. Orbital changes in the UV continuum and (non-wind)
line spectra, particularly associated with the phases before
eclipse, have also been seen in NLs and DN in outburst
\citep{mason1997,froning2001}.  The orbital effects are believed
to be caused by interactions between the disk and the mass
accretion stream from the donor star, resulting in a bulge in the
outer disk or in stream overflow, similar to the source of X-ray
dips in some X-ray binaries \citep{white1995}. Azimuthal
variations in CV winds may be tied to the disk-stream interaction
region, but the explicit link between outer disk and wind
asymmetries has not yet been discovered.

\subsubsection{Stochastic Variability}

While some CV winds show orbital modulation, others are dominated
by secular variability.  (Oddly, there is little overlap between
these two sets of systems.) \citet{prinja2000} studied high time
resolution HST/GHRS spectra of the NL, BZ~Cam.  The spectra show
strong and continuous variability in the absorption troughs, but
not the emission components, of the P~Cygni lines of CII
$\lambda1334$~\AA, CIII $\lambda1175$~\AA, Si III
$\lambda1300$~\AA, SiIV, NV, and CIV. The lines vary together and
can show optical depth changes of a factor of 5 over 1000 --
3000~km~s$^{-1}$ of the absorption troughs on 100~sec time scales.
Since all lines vary together, the variability is likely tied to
density fluctuations in the wind rather than changes in the
ionization state. In BZ~Cam, then, the picture is that of a
turbulent, stochastic wind, a contrast to the more stable winds
from luminous stars. The outflows in CVs are much more compact
than those from stars --- Prinja et al.\ note that a wind
travelling at 3000~km~s$^{-1}$ can traverse the binary separation
($\sim$50R$_{WD}$) in 2 -- 3~min --- and may therefore reflect
underlying disk fluctuations on short time scales. This chaotic,
unsteady outflow is also consistent with the behavior of the wind
in the models of \citet{proga1998}, although in other hydrodynamic
models, a more steady outflow is seen \citep{pereyra1997}.

\subsubsection{Narrow Absorption Dips}

Some CVs do not show the short time scale variability exhibited by
BZ~Cam, but do show variable line profiles in the form of narrow
absorption components to the lines.  In the NLs IX~Vel and
V3885~Sgr, very narrow ($<100$~km~s$^{-1}$), blueshifted
($-900$~km~s$^{-1}$) absorption dips appear in the absorption
troughs of several of the wind lines at times of strongest wind
activity \citep{mauche1991,hartley2002}.  Similar dips appear
intermittently in the profiles of the OVI
$\lambda\lambda$1032,1038~\AA\ doublet lines of U~Gem in outburst
\citep{froning2001}.  In U~Gem, the lines (located at $\sim
-500$~km~s$^{-1}$) appear erratically and are not correlated with
orbital phase;  sometimes, two dips appear in each of the doublet
transitions. The features are analogous to the discrete absorption
components (DACs) seen in luminous star winds, although the CV
absorption dips remain at constant velocity, while the DACs are
observed to move slowly outward with time.  Because the dips in
the CV lines do not move, many of the models invoked to explain
DACs will not apply for the CV dips.  A possible source for the
dips is the formation of localized plateaus in the outflow
velocity law, leading to pileup of ions.

\subsubsection{High Ionization Wind Lines}

Determining the formation mechanisms of winds in high accretion
rate CVs also depends on understanding cases in which the
``standard'' UV wind features do not appear.  One example is U~Gem
in outburst: with the exception of the aforementioned absorption
dips in OVI and perhaps a weak P~Cygni profile in the same, there
is no sign of an outflow in the FUV spectra of U~Gem; instead, the
lines are generally narrow, low-velocity absorption features and
sometimes are absent altogether \citep{sion1997,froning2001}.  One
possibility is that the BL luminosity plays a role in the
structure of the wind. U~Gem is one of the few CVs that shows
evidence for a luminous BL in outburst:  EUV spectra show a hot
(140,000~K) blackbody component whose luminosity, consistent with
classical accretion disk theory, is comparable to that of the
accretion disk \citep{long1996}.  The same EUV observations show
highly ionized resonant line transitions in emission.  The lines
that appear are consistent with an origin in gas photoionized by
the BL and are likely formed in a highly ionized wind.

Further understanding of the ionization structure of winds will
require comparisons of UV, EUV, and X-ray spectra in different
systems.  The lower inclination DN, SS~Cyg, has a low-luminosity
BL, for example, and shows wind features in both X-ray and FUV
spectra, but no EUV wind lines (Mauche 2004; Long et al. 2004, in
preparation).  Good fits to the X-ray spectrum can be obtained
with a wind model, but only if the individual ionizations
fractions of the lines are allowed to vary arbitrarily.  In the DN
OY~Car in superoutburst, the EUV spectrum shows strong wind
emission in resonant transitions of several ions
\citep{mauche2000}. Mauche \& Raymond note, however, that the
radiation pressure required to drive the wind in OY~Car cannot be
met by current wind models, a point that is discussed further
below.

\subsection{Limitations of the Pure Line Driving Model}

Using time-tagged HST/STIS observations of the NLs IX~Vel and
V3885~Sgr, \citet{hartley2002} tested radiatively-driven wind
models by comparing luminosity variations in the UV continuum and
in the wind lines.  For a purely radiative wind, the strength of
the wind lines should be directly correlated with the strength of
the photoionizing continuum, but they found no such correlation in
either system:  while both the continuum and the lines varied with
time, the wind had often weakened when the continuum was higher,
and two epochs with the same continuum level in IX~Vel showed
strong wind features and little to no wind emission at all,
respectively.  \citet{hartley2002} concluded that the observations
provided a serious challenge to the model of purely
radiatively-driven disk winds.

The result was not entirely surprising, as there have been other
indications that line-driving may not be the sole wind-formation
mechanism at work in CVs.  One problem that continues to plague
hydrodynamic wind models is an inability to match the observed
strengths of the wind lines. \citet{drew2000} showed that when
current determinations of mass accretion rates in luminous CVs are
assumed, the hydrodynamic models underpredict the strength of the
wind by an order of magnitude.  As a result, additional mechanical
wind drivers may be present in luminous CVs. The possible role of
centrifugal forces in driving disk winds was considered  at an
early date by \citet{cannizzo1988}, who also noted that
hydromagnetically-driven outflows would strongly influence binary
secular evolution as sinks of angular momentum. In addition,
\citet{mauche1987} showed that the presence of shocks in the wind
can cause local density enhancements that lower the overall wind
mass loss rate required to match observed wind line strengths.
Work on incorporating non-radiative forces into CV wind models is
only beginning (see Proga, this volume), but represent an
important avenue for future studies of disk winds.

\subsection{Future Work}

Most of the modeling of CV winds has concentrated on one
transition, CIV $\lambda\lambda$1548,1552~\AA.  As a result, our
understanding of the thermal and ionization structure of disk
winds remains limited.  The next steps in understanding the
formation and behavior of CV disk winds will depend on
simultaneously modeling the broad range of species and ionization
levels that appear in X-ray and UV spectra of CV winds.  To this
end, \citet{long2002} developed a Monte Carlo radiative transfer
program that, given an input WD and accretion continuum spectrum,
self-consistently calculates the thermal and ionization structure
of the wind and generates synthetic spectra at every viewing
inclination.  An example of the types of fit possible with their
program is shown in Figure~3, which shows a disk and wind model
fit to the outburst spectrum of Z~Cam, obtained with the Hopkins
Ultraviolet Telescope.  A qualitatively good fit to the observed
spectrum can be obtained with a model in which the disk has a mass
accretion rate of $6\times10^{-9}$~M$_{\odot}$~yr$^{-1}$ and the
mass loss rate in the wind is
$1\times10^{-10}$~M$_{\odot}$~yr$^{-1}$.

\begin{figure}[!ht]
\plotone{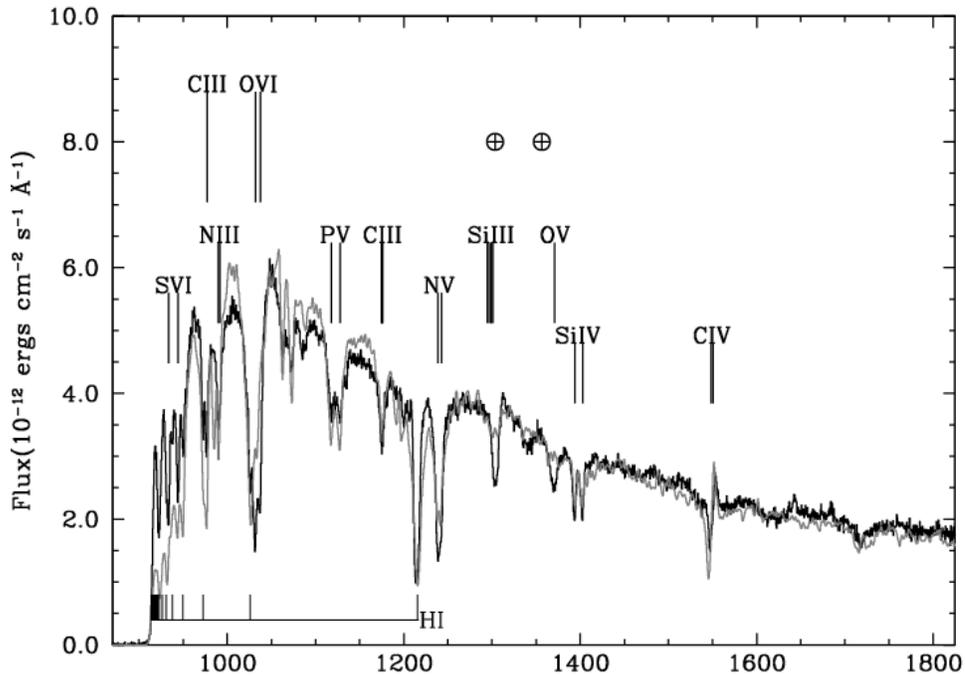} \caption{The HUT spectrum of the novalike
CV, IX Vel is shown in black.  Superimposed in gray is an
accretion disk and disk wind model of the spectrum. Figure from
Long \& Knigge 2002.}
\end{figure}

In the case of relatively simple spectrum, such as the FUV
spectrum of SS~Cyg in outburst, which is dominated by OVI (see
Figure 1), a good model fit can be achieved assuming a
steady-state accretion disk and a disk wind carrying away
$\simeq10^{-2}$ of the accreted mass \citep{froning2002}. The wind
mass loss rate obtained from the model fit
($4\times10^{-11}$~M$_{\odot}$~yr$^{-1}$) is also within a factor
of a few of the mass loss rate determined from Chandra
observations of SS~Cyg in outburst \citep{mauche2004}. Once the
spectrum becomes more complex, however, with wind signatures
appearing in many lines, as in the case of Z~Cam above, fitting
the entire spectrum and determining uniqueness of the fit
parameters becomes challenging.  CV spectra are extremely complex
and can contain emission from multiple components in the system,
parameters for some of which remain poorly constrained.
Nevertheless, the combination of high quality observations and
sophisticated models offer a promising opportunity to quantify the
structure and behavior of disk winds in CV systems.

\section{Conclusions}

Outflows, in the form of moderately collimated winds, are closely
tied to disk accretion in luminous CVs.  A combination of
observations and modeling have shown that CV winds are bipolar,
highly turbulent, and structurally complex.  Radiative
line-driving models for the source of the wind explain many of the
observations, but additional, mechanical mechanisms may play a
role in the structure and formation of the wind. Future
observations and modeling will concentrate on determining the
vertical and azimuthal distribution of the wind, its thermal and
ionization structure, and the mechanisms that influence wind
formation and behavior.

\end{document}